\documentstyle[12pt,titlepage]{article}

\newcommand{\tr}{\, {\rm tr} \,}

\newcommand{\Z}{\bf Z}

\begin{document}
\begin{titlepage}
\addtolength{\baselineskip}{.7mm}
\thispagestyle{empty}
\begin{flushright}
TIT/HEP--228 \\
NUP-A-93-13 \\
hep-th@xxx/9307116\\
July, 1993
\end{flushright}
\begin{center}
{\large{\bf
Nonlinear Renormalization Group Equation \\
for Matrix Models }} \\[15mm]
{\sc Saburo Higuchi}
\footnote{{\tt e-mail: hig@phys.titech.ac.jp}, JSPS fellow}\\[2mm]
{\it Department of Physics, Tokyo Institute of Technology, \\[1mm]
Oh-okayama, Meguro, Tokyo 152, Japan} \\[6mm]
{\sc Chigak Itoi}
\footnote{\tt e-mail: itoi@phys.titech.ac.jp} \\[2mm]
{\it Department of Physics and Atomic Energy Research Institute, \\[1mm]
College of Science and Technology, Nihon University, \\[1mm]
Kanda Surugadai, Chiyoda, Tokyo 101, Japan} \\[6mm]
{\sc Shinsuke Nishigaki}
\footnote{\tt e-mail: nsgk@phys.titech.ac.jp}
\ \ and \ \
{\sc Norisuke Sakai}
\footnote{\tt e-mail: nsakai@phys.titech.ac.jp} \\[2mm]
{\it Department of Physics,
Tokyo Institute of Technology, \\[1mm]
Oh-okayama, Meguro, Tokyo 152, Japan} \\
\vfill
{\bf Abstract}\\[5mm]
{\parbox{13cm}{\hspace{5mm}
 An exact renormalization group equation
 is derived for the free energy of matrix models.
 The renormalization group equation turns out to be nonlinear
 for matrix models, as opposed to linear for vector models.
 An algorithm for determining the critical coupling constant and
 the critical exponent is obtained.
 As concrete examples, one-matrix
  models with one and two coupling constants are analyzed and
 the exact values of the critical coupling constant and
  the associated critical exponent are found.
}}
\end{center}
\vspace*{5ex}
\end{titlepage}
\setcounter{section}{0}
\setcounter{equation}{0}
%
The matrix model provides the possibility for a nonperturbative treatment
of two-dimensional quantum gravity \cite{BIPZ}--\cite{GIMO}.
Exact solutions of the matrix model have been obtained for
two-dimensional quantum gravity
coupled to minimal conformal matter with central charge $c \le 1$.
It is important to solve matrix model representations of
two-dimensional quantum gravity
coupled to conformal matter with central charge $c > 1$.
They are interesting not only as statistical systems on a randomly
triangulated surface, but also as  string models in a target space of
arbitrary dimensions.
Although one can easily write down matrix
model candidates for cases with $c > 1$
\cite{KAZ,BRHI}, they are so far not solvable.

As an approximation scheme for obtaining critical coupling constants and
critical exponents for such unsolved matrix models,
Br\'ezin and Zinn-Justin have proposed a renormalization group
approach \cite{BRZJ}.
A similar approach has been advocated previously for
the $1/N$ expansion in various other contexts \cite{CARL}.
Several groups have examined the consequences of such an approach
\cite{ALDA},
suggesting that the result for universal quantities
(critical exponents) does not improve even if one evaluates
beta functions perturbatively up to the first few orders.
Therefore in order to make full use of such a scheme,
we first need to confirm
that the approximation method produces correct results for the
exactly solved cases.
The $O(N)$ vector model, describing a discretized
one-dimensional quantum gravity,
is known to share its double-scaling properties
with the matrix model for a discretized
two-dimensional quantum gravity \cite{VM0}.
Recently we have analyzed the vector model by means of the
renormalization group approach and
have clarified its validity and meaning \cite{HIS}.
As with the matrix models,
a perturbative evaluation of critical exponents for the
vector model does not improve up to a few orders.
Instead of a perturbative approach, we have obtained an
exact difference equation relating the free energy
for neighbouring values of $N$, with the coupling constant shifts
of order $1/N$ in infinitely many
coupling constants.
Our crucial observation was that there hold infinitely many identities
which express the freedom to reparametrize the field space.
Owing to these identities, we can re-express the flow
in the infinite dimensional coupling constant space
as an effective flow
in a finite dimensional effective coupling constant space.
Critical coupling constants and
critical exponents arise as  fixed points of the effective
beta function, and eigenvalues of the derivative matrix
of the beta function respectively.
We can understand the reason why the perturbative evaluation of
beta functions was inadequate at least in the vector model;
we have to collect all contributions from higher induced
couplings by using the reparametrization identities to
obtain a beta function.

   The purpose of this paper is to derive an exact renormalization
group equation for matrix models
and examine its validity.
Using the eigenvalue representation,
we find a difference equation relating the partition function
$Z_{N+1}(g)$ to the partition function $Z_N(g + \delta g)$
with the coupling constant shifts $\delta g_k$
of order $1/N$ in infinitely many coupling constants.
We find that the induced interactions contain terms
which are absent in usual matrix models, at least for $c \le 1$.
As with the vector model, we can derive infinitely many
identities expressing the freedom to reparametrize the matrix variables.
These identities enable us to reduce the renormalization group equation
to an effective renormalization group equation with a finite number of
coupling constants.
We observe that the effective renormalization group
equation is nonlinear in derivatives of the free energy
with respect to coupling constants.
In spite of this nonlinearity, we can provide an algorithm to compute
fixed points and critical exponents.
To illustrate this technique,
we explicitly obtain fixed points and critical exponents
together with the operator content for the
one-matrix model with one and two coupling constants.
These agree completely with the exact result.

The partition function $Z_N(g)$ of the matrix model
is defined by an
integral over an $N\times N$ hermitian matrix $\Phi$
with a generic polynomial potential $V(\Phi)$
\begin{equation}
Z_N(g)= \int d^{N^2} \Phi \exp \left[-N \tr V(\Phi) \right],
\qquad
V(\Phi)=\sum_{k=1}^{\infty}{g_k \over k}\Phi^{k}.
\end{equation}
The cubic interaction with a single coupling constant $g$
corresponds to $V(\Phi)={1 \over 2}\Phi^2+{g \over 3}\Phi^3$.
We can integrate over the angular variables to obtain an integral over
the eigenvalues $\{\lambda_j\}$ \cite{BIPZ}
\begin{equation}
Z_N(g)=c_N \int \prod^N_{i=1} d\lambda_i \ \Delta_N^2 (\{\lambda_j\})
                  \ \exp \left[-N \sum^N_{i=1} V(\lambda_i) \right],
\end{equation}
where $\Delta_N$ denotes the Van der Monde determinant
$\Delta_N(\{\lambda_j\})=\prod_{1\leq i < j\leq N}(\lambda_i-\lambda_j)$, and
$c_N=\pi^{N(N-1)/2}/\prod_{p=1}^{N} p!$ .

In order to relate $Z_{N+1}$ to $Z_{N}$, we shall integrate
the $(N+1)$-th eigenvalue $\lambda$ in $Z_{N+1}$
\begin{eqnarray}
Z_{N+1}(g) & = &
\int d^{(N+1)^2} \Phi \ {\rm e}^{-(N+1)\tr V(\Phi)} \nonumber \\
&\!\!\!=&\!\!\!
 c_{N+1} \int \prod_{i=1}^{N} d\lambda_i \ \Delta_{N}^2 (\{\lambda_j\})
 \ {\rm e}^{-(N+1) \sum_{i=1}^{N} V(\lambda_i)}
 \int d\lambda  \prod^N_{i=1} \vert \lambda  -\lambda_i \vert^2
\ {\rm e}^{-(N+1) V(\lambda )} \nonumber \\
&\!\!\!=&\!\!\! \frac{c_{N+1}}{c_N}
\int d^{N^2} \Phi \ {\rm e}^{-(N+1)\tr V(\Phi)}
\int d\lambda
\ {\rm e}^{-(N+1) V(\lambda)+2\tr \log \vert \lambda-\Phi \vert}.
\end{eqnarray}
The $\lambda$ integral can be evaluated by a saddle point
method as a power series in $1/N$
around the saddle point, since the effective potential
$(N+1)V(\lambda)-2\tr \log \vert \lambda-\Phi \vert$ is of order $O(N^1)$.
The saddle point equation
\begin{equation}
V'(\lambda^{\rm s})={2 \over N} \tr \frac{1}{\lambda^{\rm s}-\Phi} =
{2 \over N}\sum_{n=0}^{\infty}
{\tr \Phi^n \over (\lambda^{\rm s})^{n+1}}
\label{6}
\end{equation}
determines the saddle point $\lambda^{\rm s}$
as a linear combination of not only
$\tr \Phi^n$ but also nonlinear terms like
$\tr \Phi^n \tr \Phi^m$,
which are not present in the original model.
Therefore
by inserting the saddle point $\lambda^{\rm s}$ into the partition
function,
we find that infinitely many operators of the form
$\tr \Phi^m \tr \Phi^n, \ldots$ are induced
\begin{equation}
\frac{Z_{N+1}(g)}{Z_N(g)}=
 \frac{c_{N+1}}{c_{N}}
\left\langle \exp \left[ -\tr V(\Phi)
-N V(\lambda^{\rm s})+2\tr \log \vert \lambda^{\rm s}-\Phi \vert
+O(N^0) \right] \right\rangle
\label{7}
\end{equation}
{}from the renormalization transformation with respect to $N$.
Here the normalized average with respect to the measure
$ d^{N^2} \Phi \exp \left[-N\tr V(\Phi)\right]$ is denoted by
$\left\langle \ \ \ \right\rangle$.

The situation, however, simplifies if we appeal to the large-$N$ limit.
In this limit a multi-point function of $U(N)$-invariant operators
${\cal O}, {\cal O}',  \cdots$
factorizes into a product of one-point functions
\begin{equation}
\left\langle {\cal O}\ {\cal O}'  \cdots \right\rangle
= \left\langle{\cal O}\right\rangle \left\langle{\cal O}'\right\rangle\cdots
+ O\left( \frac{1}{N^2}\right) .
\label{eqn:factorization}
\end{equation}
In view of this factorization property,
eq.(\ref{7}) can be written as
\begin{equation}
\frac{Z_{N+1}(g)}{Z_N(g)}=\frac{c_{N+1}}{c_{N}}
\exp \left[ -\left\langle \tr V(\Phi) \right\rangle
- N V(\tilde \lambda)+2\left\langle
\tr\log \vert \tilde \lambda -\Phi \vert\right\rangle
+O(N^0) \right],
\label{8}
\end{equation}
where  $\tilde \lambda$ is
a function of $\left\langle \tr \Phi^m \right\rangle$
defined by averaging the saddle point equation eq.(\ref{6})
\begin{equation}
V'(\tilde \lambda)
=2\left\langle \frac{1}{N} \tr \frac{1}{\tilde \lambda-\Phi}\right\rangle
=2 \sum_{n=0}^{\infty} {1 \over \tilde \lambda^{n+1}}
\left\langle \frac{1}{N} \tr \Phi^n \right\rangle.
\label{9}
\end{equation}
Let us introduce the free energy which is normalized to vanish for
the Gaussian model
\begin{equation}
F(N,g)\equiv -\frac{1}{N^2}
\log \left[\frac{Z_N(g)}{Z_N(g_1=0,g_2=1,g_k=0~(k\ge 3))}\right],
\end{equation}
where
$Z_N(g_1=0,g_2=1,g_k=0~(k\ge 3))=2^{N/2} (\pi/N)^{N^2/2}$.
By taking the
$N\rightarrow\infty$ limit, we find the following differential
equation as a renormalization group equation
for the free energy
\begin{equation}
\left( N\frac{\partial}{\partial N} +2 \right) F(N,g)
=\left\langle \frac{1}{N} \tr V(\Phi) \right\rangle
+V( \tilde \lambda)
-2\left\langle \frac{1}{N} \tr \log \left\vert \tilde \lambda
-\Phi \right\vert\right\rangle -\frac32 +O\left( \frac1N \right).
\label{11}
\end{equation}
We observe that
this equation describes a ``flow''
in an infinite dimensional coupling constant space
which is enlarged to include interactions like $\tr \Phi^m \tr \Phi^n$.
We emphasize that this renormalization group equation does not involve a
perturbation
with respect to the coupling constants, unlike the approximation schemes
proposed in ref.\cite{BRZJ}.

We should be careful in identifying
the renormalization group flow in the coupling
constant space, since all the correlators of $\tr \Phi^m$
appearing in the right hand side of
eq.(\ref{11}) are not independent.
In fact, the partition function $Z_N(g)$ should be invariant under
reparametrization of matrix variables regular at the origin,
$\Phi \rightarrow \Phi + \varepsilon \Phi^{n+1}$ ($n\ge -1$).
Then we can construct
a tower of identities expressing this reparametrization freedom
of matrix variables, as with the case of  the vector model \cite{HIS}.
The reparametrization identities can be obtained from the usual
procedure to derive the Schwinger-Dyson equation \cite{FKN}
\begin{equation}
\int d^{N^2} \Phi \tr \frac{d}{d\Phi}  \left( \Phi^{n+1}
\exp \left[ -N\tr V(\Phi) \right] \right) =0, \ \ \ \ (n \geq -1).
\end{equation}
The reparametrization identities naturally form the Virasoro vacuum
conditions on the partition function.
After using the factorization
property (\ref{eqn:factorization}) in the large-$N$ limit,
they become
\begin{equation}
\sum_{i=0}^{n} \left\langle \frac{1}{N} \tr \Phi^i \right\rangle
\left\langle \frac{1}{N} \tr \Phi^{n-i} \right\rangle
=\left\langle \frac{1}{N} \tr\left( \Phi^{n+1} V'(\Phi) \right)\right\rangle,
\ \ (n \geq -1).
\label{14}
\end{equation}

Since the one-point function is nothing but the derivative of the
free energy with respect to coupling constants,
$
\langle\tr \Phi^k\rangle/N=k \partial F / \partial g_k ,
$
 the reparametrization identities provide infinitely many relations
among derivatives with respect to different coupling constants.

In the case of the vector model, we have shown \cite{HIS} that
one can eliminate redundant one-point functions in the right hand
side of eq.(\ref{11}) except for the first few ones corresponding
to the original potential, by applying the reparametrization
identities recursively.
To facilitate the procedure, it is convenient to
 define the expectation value of the resolvent
$ W(z)=\langle \tr (1/(z-\Phi))\rangle/N $,
with respect to which the reparametrization
identities (\ref{14}) take the form \cite{BIPZ,GIMO}
\begin{equation}
W(z)^2 -V'(z) W(z) +Q(z; V)=0,\
Q(z; V)\equiv \sum_{k \geq 1} \frac{V^{(k+1)}(z)}{k!}
\left\langle \frac{1}{N} \tr (\Phi-z)^{k-1} \right\rangle.
\end{equation}
If we are to begin with the $m$-th order potential,
$Q(z; V)$ is a polynomial in
$\langle \tr \Phi^k \rangle /N$
for $1 \leq k \leq m-2$.
Let us choose a coupling constant subspace:
$g_1=0, g_2=1, g_k=0 \ (k\ge m+1)$ with $g_3, \cdots, g_m$ arbitrary.
The first two equations of (\ref{14})
\begin{equation}
0=\left\langle \frac{1}{N} \tr \Phi \right\rangle +
\sum_{k=3}^{m} g_k \left\langle \frac{1}{N} \tr \Phi^{k-1} \right\rangle,
\quad
1=\left\langle \frac{1}{N} \tr \Phi^2 \right\rangle
+\sum_{k=3}^{m} g_k \left\langle \frac{1}{N} \tr  \Phi^{k} \right\rangle
\label{eqn:firstreparametrization}
\end{equation}
enable us to express
$\langle \tr \Phi \rangle/N$ and
$\langle \tr \Phi^2 \rangle/N$ in terms of
$\langle \tr \Phi^k \rangle/N$ for $3\leq k \leq m$.
Therefore the reparametrization identities allow us to express all of the
higher induced interactions in terms of
$\left\langle \tr \Phi^k \right\rangle/N=k\partial F/\partial g_k$
for $3 \leq k \leq m$, and in this sense the theory is renormalizable
in the large-$N$ limit.

To summarize,
the complete set of nonlinear
renormalization group equation for the one-matrix model consists of
the following three equations.
The reparametrization identities become the so-called loop
equation
\begin{equation}
 W(z)^2-V'(z) W(z) +Q\left( z ;g_3,\cdots,g_m;
 \frac{\partial F}{\partial g_3},\cdots ,
 \frac{\partial F}{\partial g_m} \right) =0 .
\end{equation}
The reparametrization identities simplify
the saddle point equation into
\begin{equation}
 V'\left( \tilde \lambda \right)^2 -4Q\left( \tilde \lambda ;g_3,\cdots ,g_m;
 \frac{\partial F}{\partial g_3},\cdots ,
 \frac{\partial F}{\partial g_m} \right) =0  .
\end{equation}
Finally the effective renormalization group equation becomes
\begin{eqnarray}
&\!\! &\!\!
\left( N\frac{\partial}{\partial N} +2 \right) F(g)
 = G \left( g, {\partial F \over \partial g} \right)
+O\left( \frac1N \right) , \nonumber \\
G\left(g,\frac{\partial F}{\partial g}\right)&\!\!\!\!\equiv&\!\!\!\!
-1- \sum^m_{k = 3}{k-2 \over 2} g_k \frac{\partial F}{\partial g_k}
+V(\tilde \lambda) -2\log \tilde \lambda
-2 \int^{\tilde \lambda}_{\pm\infty} dz \left( W(z) -\frac{1}{z} \right).
                                 \label{explicit-rge}
\end{eqnarray}
The above equations show that the renormalization group equation for
the matrix model is inevitably
nonlinear with respect to $\partial F(g)/\partial g_k$.
Nonlinearity emerges out of the renormalization group transformation
(induced interactions of the
form $\tr \Phi^m \tr \Phi^n \cdots$) and from the procedure
to eliminate redundant one-point functions.

Now we shall give an algorithm to determine fixed
points and critical exponents
as well as the free energy $F=\sum^{\infty}_{h=0} N^{-2h}
f_h(g)$.
For simplicity let us consider the case of the
 single cubic coupling for the one-matrix model.
We first restrict ourselves to the sphere topology.
The right hand side of the renormalization group equation has nonlinear
terms ($g=g_3$)
\begin{equation}
 G \left(g, {\partial F \over \partial g}\right)
=\sum_{n=0}^{\infty} \beta_n(g)
\left({\partial F \over \partial g}\right)^n.
\end{equation}
If we expand the coefficients $\beta_n$ in powers of the coupling
constant around a fixed point $g_{*}$, they are regular
\begin{equation}
\beta_n(g)=\sum_{k=0}^{\infty}\beta_{nk} (g-g_{*})^k.
\label{eqn:nonlinearbeta}
\end{equation}
The non-analyticity of the free energy should stem from solving the
differential equation (\ref{explicit-rge}).
We assume that the free energy consists of singular and regular terms
$$
F(N,g)=F^{reg}(N,g)+F^{sing}(N,g),
$$
\begin{equation}
F^{reg}=\sum_{h=0}^{\infty}N^{-2h}f_h^{reg}(g),
\quad
f_{0}^{reg}(g)=\sum_{k=0}^{\infty}a_k (g-g_{*})^k,
\end{equation}
\begin{equation}
F^{sing}=\sum_{h=0}^{\infty}N^{-2h}f_h^{sing}(g),
\quad
f_{0}^{sing}(g)=\sum_{k=0}^{\infty}b_k (g-g_{*})^{k+\gamma}.
\end{equation}
By definition, the coefficient $b_0\not=0$.
By comparing the power series expansion of the renormalization group
equation, we find
consistency conditions for
the above expansion to be valid.
The most singular term  $(g-g_{*})^{\gamma-1}$ determines
the fixed point
\begin{equation}
0 =  b_0 \gamma \sum_{n=1}^{\infty} \beta_{n0} n a_1^{n-1}.
\label{eqn:fixedpoint}
\end{equation}
The next singular term  $(g-g_{*})^{\gamma}$ determines
the critical exponent $\gamma$ since $b_0\not=0$
\begin{equation}
2b_0  =  b_0 \gamma \left[
\sum_{n=1}^{\infty} \beta_{n1} n a_1^{n-1}
+2a_2 \sum_{n=2}^{\infty} \beta_{n0} n(n-1) a_1^{n-2} \right].
\label{eqn:criticalexponent}
\end{equation}
They contain two coefficients $a_1$ and $a_2$ of the regular
part.
However, the consistency conditions for
terms of order $(g-g_{*})^1$ and $(g-g_{*})^2$
determine  $a_1$ and $a_2$ respectively
\begin{equation}
2a_1= \sum_{n=0}^{\infty} \beta_{n1} a_1^{n},\
2a_2= \sum_{n=0}^{\infty} \beta_{n2} a_1^{n}
+2a_2 \sum_{n=1}^{\infty} \beta_{n1} n a_1^{n-1}
+2{a_2}^2 \sum_{n=2}^{\infty} \beta_{n0} n(n-1) a_1^{n-2}.
\label{eqn:analyticcoeff}
\end{equation}
In terms of $G$, eqs.\
(\ref{eqn:fixedpoint})--(\ref{eqn:analyticcoeff})
are equivalent to
\begin{eqnarray}
\label{eqn:1parameter}
\!\!\!\!\!\!0=G_{,a_1}(g_* ,a_1)\!\!\!\!\!&,&\!\!
\frac2\gamma =G_{,ga_1}(g_* ,a_1)+2a_2 \ G_{,a_1 a_1} (g_*, a_1)\ , \\
\!\!\!\!\!\!2a_1=G_{,g}(g_*, a_1)\!\!\!\!&,&\!\!
2a_2=\frac{1}{2} G_{,gg}(g_*, a_1)+2a_2 \ G_{,ga_1}(g_*, a_1)
+ 2 {a_2}^2 \ G_{,a_1 a_1} (g_*, a_1) .\nonumber
\end{eqnarray}
By solving these four equations,
we can determine four quantities, namely
the fixed point $g_*$,
the coefficient $a_1$,
the critical exponent $\gamma$,
and the coefficient $a_2$.
The susceptibility exponent $\gamma_0$ for the sphere
topology is related to $\gamma$ via $\gamma=2-\gamma_0$.

In the one coupling case at hand,
$G(g,a_1)$ is given by
\begin{eqnarray}
&\!& G(g,a_1)=-1-\frac{g}{2} a_1 +\frac{1}{2} \tilde{\lambda}^2
 +\frac{g}{3} \tilde{\lambda}^3 -2\log \tilde{\lambda} \nonumber \\
&\!& -\int^{\tilde{\lambda}}_{\pm\infty} dz \left[ z+gz^2-
\sqrt{\left(z+gz^2 \right)^2
  -4 \left( 1+gz-g^2 +3g^3 a_1 \right)}-\frac{2}{z} \right]
\end{eqnarray}
with
$\tilde{\lambda}=\tilde{\lambda} (g,a_1)$
being the zero of the square root.
We have found that
the set of equations (\ref{eqn:1parameter}) have three solutions;
two of them are the nontrivial ultraviolet fixed point
which agree with the exact result for the $m=2$ critical points
\begin{equation}
g_{*} = \pm (432)^{-1/4},\ \
2-\gamma_0 = 5/2,\ \
a_1 = \pm 3^{-1/4} 2 ( 1 - \sqrt{3}),\ \
a_2 = -4(17\sqrt{3} + 30).
  \label{m=2exactsol}
\end{equation}
The other fixed point is located at the origin: $g_{*}=0$ and
$2-\gamma_0=-4$ which agrees with the na\"{\i}ve scaling as expected.
At each fixed point, the other consistency conditions determine all
the other coefficients
$a_k$ $(k\ge 0)$ and $b_k/b_0$ ($k\ge 1$) recursively except for the
overall normalization of the singular term $b_0$.
Although the equation for $a_2$ is quadratic,
 we choose the branch which continues to the unique solution at the origin
(the Gaussian fixed point).
In this way we can obtain the series expansion of the sphere free energy
around the fixed point, up to one integration constant.
This situation is completely analogous to the vector model case.

Up to now we have implicitly assumed that the exponent $\gamma$
is an irrational number.
In the case of the rational exponent as in eq.(\ref{m=2exactsol}),
the singular part is still well
defined, but its higher power will contribute to terms of integer
powers of \mbox{$(g-g_*)$ }
\footnote{
  Since the distinction between the regular and the singular
  parts of the free energy becomes meaningless for integer $\gamma$,
  we assume that $\gamma$ is not an integer.
  }.
Therefore the equation to determine the regular part acquires a
contribution from the singular part at higher orders
and hence equations to determine
the regular and singular parts mix each other.
This mixing does not occur in eqs.(\ref{eqn:analyticcoeff})
which determine the coefficients $a_1$ and $a_2$
at least for the solutions (\ref{m=2exactsol}).
Since susceptibility exponents for $c \le 1$ are known to be
$\gamma \ge 2$ for exact solutions,
eqs.(\ref{eqn:analyticcoeff}) are also valid in these $ c \le 1$ cases.

It is straightforward to generalize the algorithm to the case of many
coupling constants.
The expansion around the fixed point $g_*$ is given in terms of
$ \Delta g_i \equiv g_i -g_{i*}$
\begin{eqnarray}
f_0^{reg} (g) &=&
\sum^{\infty}_{n=0} a^{i_1 \cdots i_n}(g_*) \Delta g_{i_1} \cdots
\Delta g_{i_n},
\nonumber\\
f_{0}^{sing} (g) &=&
\sum_i \left(
{V_i}^j \Delta g_j
\right)^{\gamma_{(i)}}
\sum^{\infty}_{n=0}
b^{i_1 \cdots i_n}(g_*) \Delta g_{i_1} \cdots \Delta g_{i_n},
\label{f0rgw}
\end{eqnarray}
where we have used a transformation matrix ${V_i}^j $ which
diagonalizes  the exponent matrix.
We obtain by expanding the renormalization group equation in
powers of $\Delta g$
\begin{eqnarray}
0=\frac{\partial G}{\partial a^i} &,&
\sum_k {(V^{-1})_j} ^k \frac{2}{\gamma_{(k)}} {V_k}^i
=\frac{\partial^2 G}{\partial g_i \partial a^j}
+2 a^{ik} \frac{\partial^2 G}{\partial a^j \partial a^k} ,
\nonumber\\
2 a^i=\frac{\partial G}{\partial g_i} &,&
2 a^{ij}=\frac{1}{2} \frac{\partial^2 G}{\partial g_i \partial g_j}
+a^{ik} \frac{\partial^2 G}{\partial g_j \partial a^ k}
+a^{jk} \frac{\partial^2 G}{\partial g_i \partial a^ k}
+2 a^{ik} a^{j\ell} \frac{\partial^2 G}{\partial a^k \partial a^\ell} ,
\label{eqn:2parameters}
\end{eqnarray}
where they come from terms of order
$O \left( ({V_i}^j \Delta g_j)^{\gamma_{(i)}-1} \right)$,
$O\left( ({V_i}^k \Delta g_k)^{\gamma_{(i)}-1} ({V_j}^k \Delta g_k ) \right)$,
$O(\Delta g_i)$ and $O(\Delta g_i \Delta g_j)$,
respectively.
These four form a closed set of equations to determine fixed points
and critical exponents  as in the one coupling case.

As an illuminating
 example, we analyze the case of two couplings ($g_3$ and $g_4$)
for the
one-matrix model.
We find five solutions for the set of equations
(\ref{eqn:2parameters}).
Firstly, we find fixed points at
$(g_{3*},g_{4*}) = (\pm 0.3066\ldots,0.02532\ldots)$.
The free energy behaves in their neighbourhoods as
\begin{equation}
F^{sing}(N,g_3,g_4)
\sim b\cdot [\Delta g_3 \pm (-4.472\ldots)\cdot\Delta g_4 ]^{7/3}
+ b'\cdot [\Delta g_3 \pm (-6.209\ldots)\cdot\Delta g_4 ]^{7/2}.
\label{2coupm=3}
\end{equation}
The positions of the critical points and the transformation matrices $V_i^j$
in eq.(\ref{f0rgw}) can be expressed as roots of algebraic equations,
though we present them in numerical forms for brevity
in eq.(\ref{2coupm=3}).
We find two singular terms with the critical exponents
$\gamma= 7/3 $ and $7/2$ at this fixed point.
They correspond to the gravitational dressing of operators
with bare conformal dimension $\Delta_0 = -1/5$ and $0$,
respectively \cite{DDK}.
This fixed point should describe the Lee-Yang edge singularity.
The fixed point and the critical exponents agree with the exact
result for the $m=3$ critical point.
We also have the $m=2$ fixed points which we have just seen in the
one coupling constant case eq.(\ref{m=2exactsol}),
$
(g_{3*},g_{4*}) = (\pm (432)^{-1/4},0).
$
The critical behaviors around these points are
\begin{equation}
F^{sing}(N,  g_3 ,  g_4)
 \sim b\cdot
     \left[ \Delta g_3 \pm
       \left(  \frac{3^{1/4}}{4} - \frac{3^{7/4}}{2}
       \right)
      \Delta g_4
     \right]^{5/2}
+ b'\cdot [\Delta g_4]^{-6}.
\label{2coupm=2}
\end{equation}
The first  critical exponent $5/2$ corresponds to
the gravitationally dressed cosmological term for the $m=2$ critical point.
We find no other positive exponent,
in agreement with the fact that there is no primary field other
than the identity in the matter conformal field theory before
gravitational dressing.
The last one is the trivial fixed point at $(g_{3*}, g_{4*})=(0, 0)$
which gives the na\"{\i}ve scaling behavior
\begin{equation}
F^{sing}(N, g_3 , g_4)
 \sim  b \cdot [\Delta g_4]^{-2} + b' \cdot [\Delta g_3]^{-4} .
\label{2coupm=1}
\end{equation}

In the exact solution it is known that there are the $m=3$ critical
points and the $m=2$ critical lines in the two dimensional coupling
constant space.
We observe that the critical behavior of the second term
in eqs. (\ref{2coupm=3})--(\ref{2coupm=1}) is realized at each
fixed point  when one approaches the fixed point along the $m=2$
critical line.
It is interesting that all of the above solutions that we
find turn out to be on the $m=2$ critical lines.
However, let us note that not every point on the $m=2$ critical lines
is a fixed point, as with the case of the vector model.

If we compare our result with the exact one,
another solution corresponding to the $m=2$ critical
point $ (g_{3*},g_{4*}) = ( 0, -1/12)$ is expected to be present.
It is likely that such a solution exists,  though we have not proved it
due to the subtlety arising from the ${\Z}_2$-invariance of the critical
potential.

Next we turn to the scaling behavior of the higher genus contributions
$f_h(g)$ to the free energy $F=\sum^{\infty}_{h=0} N^{-2h} f_h(g)$.
Again for brevity we consider the single coupling case.
We can start from the difference equation and take into account the
shift $\delta g_k$ of higher orders in $1/N$.
A differential equation with additional terms
can be derived from the difference equation
by retaining the higher order terms in expanding in $1/N$.
If we expand the free energy into $1/N$ series,
we can separate the partial differential equation into a set of ordinary
differential equations for each genus contribution.
It is important to realize that the additional contributions
introduced into the right hand side $G$ of the renormalization group
equation carry additional powers of $1/N$.
Therefore they will not appear in the coefficient $\beta_n(g)$
\begin{equation}
\left( N\frac{\partial}{\partial N} + 2 \right) F(N,g) =
 r(g,N)
+\sum_{h=0}^{\infty} N^{-2h} \sum_{n=1}^{\infty}
 \beta_n(g) \sum_{%
        \begin{scriptsize}
        \begin{array}{c}
          h_1+\cdots+h_n=h,\\
          h_j\geq 0\\
        \end{array}%
        \end{scriptsize}
}
 \prod_{j=1}^{n} \frac{d f_{h_j}}{dg}.
\end{equation}
The inhomogeneous term $r(g,N)$ does get various higher order
contributions including those terms from $f_{h'}$ for $h' \le h-1$.
Therefore we obtain
\begin{equation}
2(1-h)f_h(g)=r_h(g)
+\sum_{n=1}^{\infty} \beta_n(g) \sum_{%
        \begin{scriptsize}
        \begin{array}{c}
          h_1+\cdots+h_n=h,\\
          h_j\geq 0\\
        \end{array}%
        \end{scriptsize}
}
 \prod_{j=1}^{n} \frac{d f_{h_j}}{dg},
\end{equation}
where $r_h(g)$ is defined by $r(g, N)=\sum N^{-2h} r_h(g)$.
Only the inhomogeneous term depends on genus, whereas the
beta functions $\beta_n(g)$ for $n\ge 1$ are universal
for any genus.
Once we realize this structure of the renormalization group equation,
we can repeat the same argument as in the case of the sphere to obtain
the scaling behavior of the higher genus contributions.
We find the fixed point condition is the same as
eq.(\ref{eqn:fixedpoint}) except that the coefficient $b_0$ is now
replaced by $b_0^h$ of the singular term of the genus $h$ free energy.
The critical exponent condition becomes
\begin{equation}
2(1-h)b_0^h  =  b_0^h \gamma^h \left[
\sum_{n=1}^{\infty} \beta_{n1} n a_1^{n-1}
+2a_2 \sum_{n=2}^{\infty} \beta_{n0} n(n-1) a_1^{n-2} \right].
\label{eqn:hcriticalexponent}
\end{equation}
We find that
the coefficients $a_1$ and $a_2$ of
the regular term are needed to fix these equations.
However, it is notable that they are precisely those
regular terms for the {\it sphere} free energy.
Therefore the same conditions as the sphere case
(\ref{eqn:analyticcoeff})
are sufficient to determine them.
In this way we find that the fixed point is universal for any genus and
the critical exponent $\gamma^h$ for genus $h$ is given by
\begin{equation}
\gamma^h=(1-h)\gamma_1, \qquad \gamma_1+\gamma_0=2.
\end{equation}
This result explains the double scaling behavior for the singular
part of the free energy
\begin{eqnarray}
F^{sing}(N, g)
 &\!\!\! = &\!\!\!
\sum_{h=0}^{\infty} N^{-2h}f_h^{sing}(g)
=\sum_{h=0}^{\infty} N^{-2h}
(g - g_{*})^{2 - \gamma_0-\gamma_1 h} b_0^h \nonumber\\
&\!\!\! = &\!\!\! (g - g_{*})^{2- \gamma_0}
f^{sing}\bigl(N^{2/\gamma_1}(g-g_{*})\bigr).
\end{eqnarray}

So far we have employed the eigenvalue representation of the
matrix integral and have attempted to integrate the $(N+1)$-th
eigenvalue.
We can, however, also use the coset technique
proposed in ref.\cite{BRZJ} where one integrates over the
$(N+1)$-th row and column vector
$\Phi_{N+1\;j}=\Phi_{j\;N+1}^{\dagger}, ~ ( 1 \le j \le N)$ and the singlet
 $\Phi_{N+1\;N+1}\equiv \alpha$
retaining only the $N \times N$ matrix
$\Phi_{ij},~ ( 1 \le i, j \le N)$.
We can combine their technique with the reparametrization identities.
Therefore we obtain a nonlinear renormalization group equation in the case
of a single coupling constant
\begin{eqnarray}
&&\left(N \frac{\partial}{\partial N} + 2 \right) F(N,g)
 = \tilde G \left(g, \frac{\partial F}{\partial g}\right)
+ O\left(\frac{1}{N}\right), \\
&&\tilde G\left(g,\frac{\partial F}{\partial g}\right)  \equiv
   - \frac{g}{2}  \frac{\partial F}{\partial g}
   + \frac{\tilde{\alpha}^2}{2}
   + \frac{g \tilde{\alpha}^3 }{3}
   + \log ( 1 + g \tilde{\alpha})
  + \int^{-\frac{1+g \tilde{\alpha}}{g}}_{\mp\infty} dz
                                \left(W(z) - \frac{1}{z}\right), \nonumber\\
  \label{coset-rge}
\end{eqnarray}
where the saddle point $\tilde{\alpha}$ of $\alpha \equiv \Phi_{N+1\;N+1}$
is determined using the expectation value of the resolvent
\begin{equation}
  \tilde{\alpha} + g \tilde{\alpha}^2 - W\left(-\frac{1+g
    \tilde{\alpha}}{g}\right) = 0.
\end{equation}
By using the same power series expansion method, we have computed
the fixed point and the critical exponents from these equations too.
We get the same result as with the eigenvalue method.
Therefore we believe their method is equivalent to ours in the
eigenvalue representation.

To obtain the expression (\ref{coset-rge}),
we have evaluated the integration
over $\Phi_{N+1\;N+1} \equiv \alpha$ by the saddle point method.
Let us note that it is essential to perform the
integration over $\alpha$ to reproduce the exact
results.  It is not correct to ignore the degree of freedom
corresponding to $\alpha$ by the large-$N$ assumption.

The disadvantage of the coset technique compared to ours is that one
has to introduce auxiliary fields
when interactions like $(\Phi_{N+1\;i}\;\Phi_{i\;N+1})^2$ are present
as in the case of quartic or higher degree potentials.
On the other hand, our eigenvalue representation
allows all the potentials to be treated on the same
footing.
Nevertheless,
the coset technique does seem to have wider applicability,
especially if we try to consider the cases unsolved so far.
Therefore we would like to consider this method as a possible convenient
technique for examining matrix models such as those
corresponding to $c > 1$.

{}From our study we have now learned that the linearized
approximation neglecting the nonlinear terms in our renormalization
group equation provides a result not too far from the exact result, at
least in the cases which we have studied.  Therefore we hope to employ the
linearized approximation to study models admitting no exact solution.
This is equivalent to ignoring higher order terms with respect to the
coupling constant $g$ in the right hand side $G$ of the
renormalization group equations. We stress again that this
approximation should be done only after reparametrization identities
are taken into account. We have already computed a number of such
cases.

The nonlinearity of the
renormalization group equations also makes it difficult
to draw the
renormalization group flow or to characterize the $m=2$ critical
line as a trajectory of the flow.
However, once we make a linearized approximation,
it is easy to know the flow vector field in the coupling constant space.
Therefore we can use the linearized renormalization group equation to
visualize the renormalization group flow approximately or
qualitatively. This approximation should be good near the origin.
We are currently trying to find a scheme for a more precise
study of the renormalization group flow.

It is straightforward to write down the renormalization group equation
and the saddle point equation for any type of matrix model.
It would be nice to reduce the general multi-matrix
reparametrization identities (loop equations)
to the more tractable form of a single matrix case.
We have already observed that our method can be extended
in a straightforward way  to deal with the two-matrix model,
which is known to describe any minimal matter coupled to
two-dimensional gravity.
We can also find operator content at any fixed points in our
approach,
as in eqs. (\ref{2coupm=3}) -- (\ref{2coupm=1}) of the one-matrix case.
We hope to report the results in subsequent publications.

Note added:
After we finish writing up this paper, we noticed that a preprint
\cite{ITOH} has appeared. In the paper, he analyzed the one-matrix
model with the cubic potential using the coset technique.  He used the
first several reparametrization identities, but has not been able to
take account of contributions from all higher induced interactions
which are of the same order. He has not included the $\alpha$-integral.

The authors thank P.~Crehan for a careful reading of the manuscript.
One of the authors (S.H.) thanks J. Zinn-Justin, I.~K.~Kostov, E. Br\'ezin,
and P.~Di~Vecchia, for discussion and hospitality during his visit to
Saclay, Ecole Normale, and Nordita.
Another one (C.I.) thanks M.~Siino for helping him use
Mathematica at the early stage of this work.
This work is supported in part by Grant-in-Aid for Scientific
Research (S.H.) and (No.05640334)(N.S.), and Grant-in-Aid for Scientific
Research for Priority Areas (No.05230019)(N.S.) {}from
the Ministry of Education, Science and Culture.

\end{document}